\documentclass[pdflatex,sn-mathphys-num]{sn-jnl}% Math and Physical Sciences Numbered Reference Style
\usepackage{graphicx}%
\usepackage{multirow}%
\usepackage{amsmath,amssymb,amsfonts}%
\usepackage{amsthm}%
\usepackage{mathrsfs}%
\usepackage[title]{appendix}%
\usepackage{xcolor}%
\usepackage{textcomp}%
\usepackage{manyfoot}%
\usepackage{booktabs}%
\usepackage{algorithm}%
\usepackage{algorithmicx}%
\usepackage{algpseudocode}%
\usepackage{listings}%
\usepackage{array}
\usepackage{dsfont}
\usepackage{makecell}
\theoremstyle{thmstyleone}%
\theoremstyle{thmstyletwo}%

\theoremstyle{thmstylethree}%

\ifx \bisbn   \undefined \def \bisbn  #1{ISBN #1}\fi
\ifx \binits  \undefined \def \binits#1{#1}\fi
\ifx \bauthor  \undefined \def \bauthor#1{#1}\fi
\ifx \batitle  \undefined \def \batitle#1{#1}\fi
\ifx \bjtitle  \undefined \def \bjtitle#1{#1}\fi
\ifx \bvolume  \undefined \def \bvolume#1{\textbf{#1}}\fi
\ifx \byear  \undefined \def \byear#1{#1}\fi
\ifx \bissue  \undefined \def \bissue#1{#1}\fi
\ifx \bfpage  \undefined \def \bfpage#1{#1}\fi
\ifx \blpage  \undefined \def \blpage #1{#1}\fi
\ifx \burl  \undefined \def \burl#1{\textsf{#1}}\fi
\ifx \doiurl  \undefined \def \doiurl#1{\url{https://doi.org/#1}}\fi
\ifx \betal  \undefined \def \betal{\textit{et al.}}\fi
\ifx \binstitute  \undefined \def \binstitute#1{#1}\fi
\ifx \binstitutionaled  \undefined \def \binstitutionaled#1{#1}\fi
\ifx \bctitle  \undefined \def \bctitle#1{#1}\fi
\ifx \beditor  \undefined \def \beditor#1{#1}\fi
\ifx \bpublisher  \undefined \def \bpublisher#1{#1}\fi
\ifx \bbtitle  \undefined \def \bbtitle#1{#1}\fi
\ifx \bedition  \undefined \def \bedition#1{#1}\fi
\ifx \bseriesno  \undefined \def \bseriesno#1{#1}\fi
\ifx \blocation  \undefined \def \blocation#1{#1}\fi
\ifx \bsertitle  \undefined \def \bsertitle#1{#1}\fi
\ifx \bsnm \undefined \def \bsnm#1{#1}\fi
\ifx \bsuffix \undefined \def \bsuffix#1{#1}\fi
\ifx \bparticle \undefined \def \bparticle#1{#1}\fi
\ifx \barticle \undefined \def \barticle#1{#1}\fi
\bibcommenthead
\ifx \bconfdate \undefined \def \bconfdate #1{#1}\fi
\ifx \botherref \undefined \def \botherref #1{#1}\fi
\ifx \url \undefined \def \url#1{\textsf{#1}}\fi
\ifx \bchapter \undefined \def \bchapter#1{#1}\fi
\ifx \bbook \undefined \def \bbook#1{#1}\fi
\ifx \bcomment \undefined \def \bcomment#1{#1}\fi
\ifx \oauthor \undefined \def \oauthor#1{#1}\fi
\ifx \citeauthoryear \undefined \def \citeauthoryear#1{#1}\fi
\ifx \endbibitem  \undefined \def \endbibitem {}\fi
\ifx \bconflocation  \undefined \def \bconflocation#1{#1}\fi
\ifx \arxivurl  \undefined \def \arxivurl#1{\textsf{#1}}\fi
\csname PreBibitemsHook\endcsname

\begin{document}

\title[Article Title]{Fault-tolerant and secure long-distance quantum communication via uncorrectable-error-injection}

\author*[1]{\fnm{IlKwon} \sur{Sohn}}\email{d2estiny@kisti.re.kr}

\author[1]{\fnm{Boseon} \sur{Kim}}\email{boseon12@kisti.re.kr}
\equalcont{These authors contributed equally to this work.}

\author[1]{\fnm{Kwangil} \sur{Bae}}\email{kibae@kisti.re.kr}
\equalcont{These authors contributed equally to this work.}

\author[1]{\fnm{Wooyeong} \sur{Song}}\email{wysong@kisti.re.kr}
\equalcont{These authors contributed equally to this work.}

\author[1]{\fnm{Chankyun} \sur{Lee}}\email{chankyunlee@kisti.re.kr}
\equalcont{These authors contributed equally to this work.}

\author[2,3]{\fnm{Kabgyun} \sur{Jeong}}\email{kgjeong6@snu.ac.kr}
\equalcont{These authors contributed equally to this work.}

\author[1]{\fnm{Wonhyuk} \sur{Lee}}\email{livezone@kisti.re.kr}
\equalcont{These authors contributed equally to this work.}

\affil[1]{\orgdiv{Quantum Network Research center}, \orgname{Korea Institute of Science and Technology Information}, \orgaddress{\city{Daejeon}, \postcode{34141},\country{Republic of Korea}}}

\affil[2]{\orgdiv{Research Institute of Mathematics}, \orgname{Seoul National University}, \orgaddress{\city{Seoul}, \postcode{08826}, \country{Republic of Korea}}}

\affil[3]{\orgdiv{School of Computational Sciences}, \orgname{Korea Institute for Advanced Study}, \orgaddress{\city{Seoul}, \postcode{02455}, \country{Republic of Korea}}}

\abstract{
Quantum networks aim to facilitate the fault-tolerant and secure transmission of quantum states across distant devices.
The widely adopted quantum teleportation scheme requires multiple rounds of entanglement swapping and purification, leading to significant resource overhead and operational complexity.
In this study, we propose a novel fault-tolerant and secure quantum communication scheme based on uncorrectable error injection.
Our method exploits a quantum state encoding scheme based on quantum error correction codes, which strategically introduces uncorrectable errors to enhance security.
It eliminates the need for entanglement distribution while reducing resource requirements.
The injected errors protect against eavesdropping by preventing unauthorized parties from retrieving meaningful information.
Security analysis shows that as the data length and encoded message size increase, information leakage becomes negligible relative to the size of the total message.
Comparative performance analysis with existing approaches indicates that our method reduces transmission overhead while maintaining comparable fidelity in low-error regimes.
These findings suggest that the proposed method offers a scalable and practical alternative for secure long-distance quantum communication, distributed quantum computing, and future quantum internet applications.
}



\maketitle

\section{Introduction}
Quantum networks provide a framework for distributing quantum information across physically separated quantum processors, utilizing quantum entanglement~\cite{qi00, qi01, qi02, qi03, qi04}.
To enable secure and efficient quantum communication, it is essential to reliably transmit arbitrary quantum states over long distances.

Quantum teleportation provides the foundation for various quantum communication tasks and serves as a fundamental protocol that utilizes quantum entanglement, as well as quantum and classical links, to transmit arbitrary quantum states~\cite{qt00, qt01, qt02}.
However, it requires a pre-shared entangled pair between the sender and receiver.
If the distance is longer than the maximal achievable distance for a single distribution link, entanglement swapping is necessary to extend the distribution range~\cite{es00, es01}.
Additionally, to enhance the fidelity of shared entanglement, entanglement purification must be performed~\cite{ep00, ep01, ep02, ep03, ep04, ep05, ep06}, as reported in~\cite{lded00}.

\begin{figure}[ht!]
\centerline{\includegraphics[width=.5\linewidth]{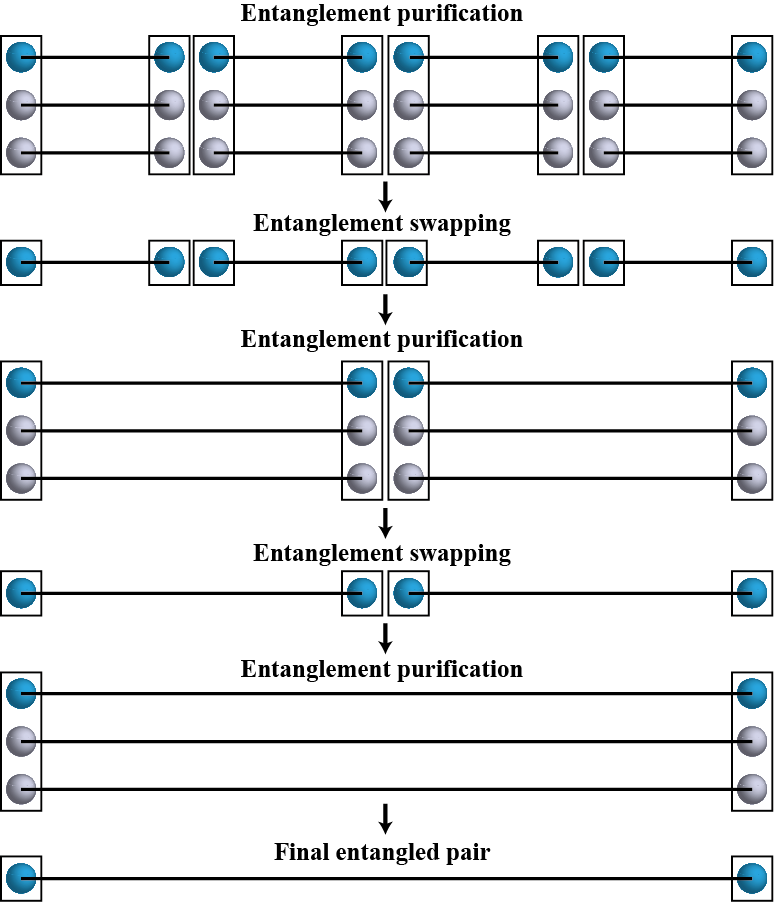}}
\caption{\textbf{Schematic of the long-distance entanglement distribution} Blue spheres represent the entangled pairs and gray spheres represent the ancilla qubits used for purification.
By repeatedly performing entanglement purification to enhance fidelity and entanglement swapping, entangled pairs can be shared over long distances.}
\label{fig:lded}
\end{figure}

In a linear optical setup, entanglement swapping has a success probability of only 50$\%$ due to the limitations of physical Bell state measurement (BSM), causing the success probability to decrease exponentially with the number of nodes over long distances.
To mitigate this, logical BSM using quantum error correction codes (QECCs) can be performed, increasing the success probability to $1-1/2^{n_{bsm}}$ based on the code length $n_{bsm}$, albeit introducing an $n_{bsm}$-fold overhead~\cite{lbsm00, lbsm01}. The overhead further increases when purification is considered.
Using the purification method described in~\cite{ep06}, high fidelity can be achieved with only two ancillary qubits per entangled pair (e.g., approximately 0.995 fidelity with a physical error rate of 0.1).
However, when the distance between the sender and receiver is significant and requires multiple stages of entanglement swapping and purification, as illustrated in figure~\ref{fig:lded}, the overhead increases exponentially with the number of relay nodes.

Given these challenges, if overhead is inevitable and qubit transmission is necessary for entanglement distribution, an alternative approach is to encode the transmitted information using QECCs and transmit it in a manner similar to classical communication~\cite{qec00, qec01, qec02, qec03, qec04, qec05, qec06, qec07, qec08, qc18}.
Nevertheless, quantum teleportation inherently possesses a degree of security because entangled pairs do not contain any information about the transmitted quantum states~\cite{qt02}.
Thus, for a fair comparison, security must also be ensured when transmitting encoded quantum states.

In this study, we introduce a novel scheme that encodes quantum states with QECCs and injects uncorrectable errors to enable fault-tolerant and secure long-distance transmission of arbitrary quantum states.
Our approach addresses the limitations of current quantum state transmission methods and provides a scalable solution for quantum communication.

\section{Uncorrectable-error-injection-based fault-tolerant and secure quantum communication}
\begin{figure*}[t]
\centering
\includegraphics[width=1\linewidth]{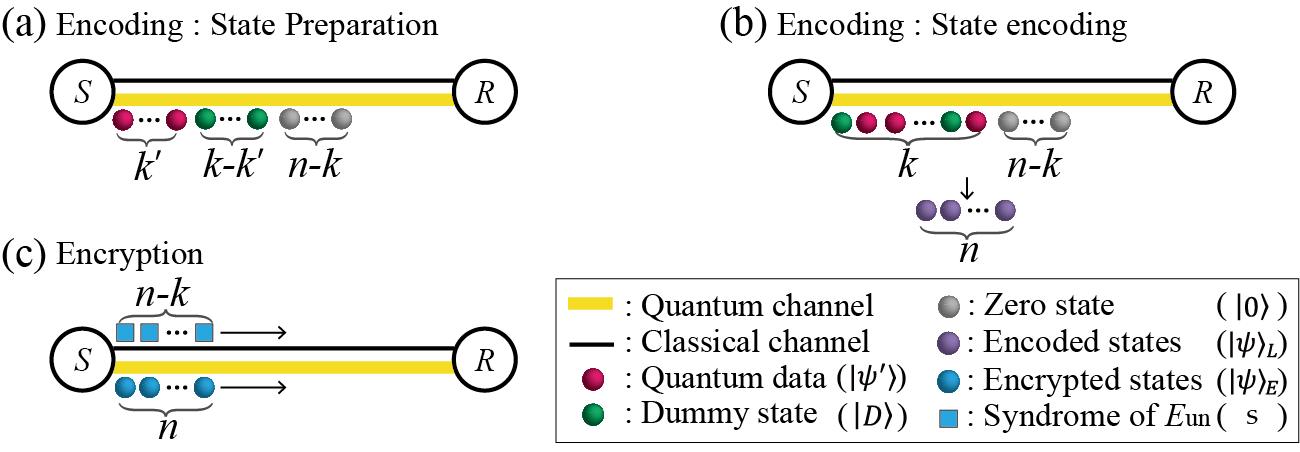}
\caption{\textbf{Schematic of the proposed scheme's encoding, and encryption.}
\textbf(a) The sender prepares an arbitrary quantum data $|\psi'\rangle = \sum_{i=0}^{2^{k'}-1} c_i |m_i\rangle$, which is to be transmitted. We assume that the quantum channel is noisy and insecure, while the classical channel is error-free and authenticated.
This state is embedded into a larger quantum register by appending mutually unbiased dummy states $|D\rangle= \bigotimes_{i=1}^{k-k'} X^{\mathbf{\kappa^{1}_{i}}}Z^{\mathbf{\kappa^{2}_{i}}}H^{\mathbf{\kappa^{3}_{i}}} |0\rangle^{\otimes k-k'}$ and zero ancillary qubits, yielding $|\psi\rangle = |\psi'\rangle|D\rangle|0\rangle^{\otimes n-k}$.
(b) The $k'$ quantum data qubits and $k-k'$ dummy states are randomly shuffled into $k$ slots.
This permutation is implemented by the operator $P_{\mathbf{\kappa^4}}$, which is determined by a random $k$-bit key $\mathbf{\kappa^4}$ with a Hamming weight of $k'$.
This state is expressed as $|\psi\rangle = U_c |\psi'\rangle |0\rangle^{\otimes n-k'}$, where $U_c = (P_{\kappa^4} \otimes I^{\otimes n-k})(I^{\otimes k'} \otimes U_{MUB} \otimes I^{\otimes n-k})$ denotes the composite operation including permutation and mutually unbiased basis transformation.
This state is subsequently encoded using a $[[n,k,d]]$ QECC, yielding the logical state $|\psi\rangle_L = U_E |\psi\rangle$.
(c) An uncorrectable error $E_{un}$ is intentionally injected into the logical state to generate the encrypted state $|\psi\rangle_E = E_{un}|\psi\rangle_L$.
The sender transmits both the syndrome $s$ corresponding to $E_{un}$ and the encrypted state $|\psi\rangle_E$ to the receiver, who performs syndrome-based decoding and verification to recover the original message.}
\label{fig:pro1}
\end{figure*}

\begin{figure*}[t]
\centering
\includegraphics[width=1\linewidth]{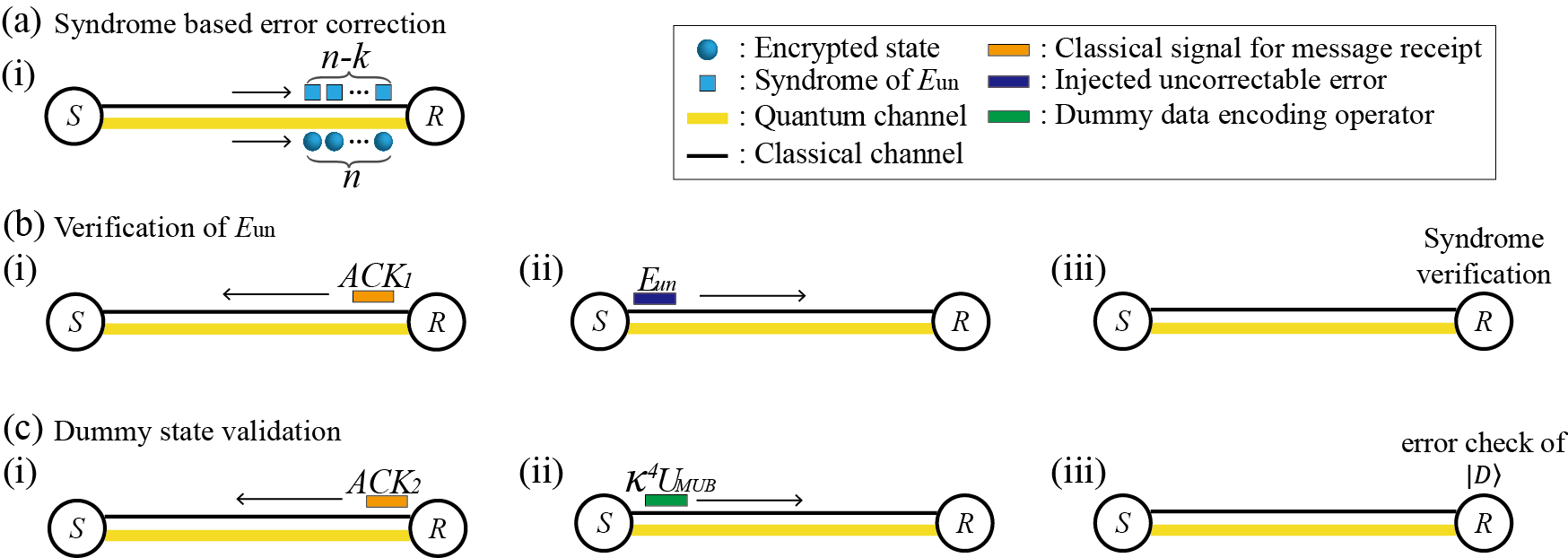}
\caption{\textbf{Schematic of the proposed scheme's decryption process}:
(a-i) The receiver performs syndrome-based error correction on the received encrypted state using the attached syndrome.
(b-i) The receiver transmits $\textit{ACK}_1$ to notify successful receipt of the state.
(b-ii) Upon receiving the acknowledgment, the sender transmits the injected uncorrectable error $E_{un}$ to the receiver.
(b-iii) The receiver applies $E_{un}$ and verifies whether an all-zero syndrome is extracted; the protocol is aborted if the syndrome is not a zero vector.
(c-i) If verification succeeds, the receiver sends $\textit{ACK}_2$ to the sender.
(c-ii) The sender then transmits $P_{\kappa^4}$ and $U_{MUB}$.
(c-iii) The receiver uses them to validate the dummy states $|D\rangle$ and check for possible eavesdropping attempts.}
\label{fig:pro2}
\end{figure*}

In this section, we describe a scheme for transmitting quantum states in a fault-tolerant and secure manner by injecting uncorrectable errors into the encoded states.
As illustrated in figure~\ref{fig:pro1} and figure~\ref{fig:pro2}, our proposed scheme consists of four major steps: (1) encoding the quantum state using a quantum error correction code (QECC), (2) strategically injecting uncorrectable errors to enhance security and fault tolerance, (3) transmitting the encoded state through a noisy and insecure quantum channel, and (4) performing syndrome-based error correction at the receiver to recover the original quantum state while verifying its authenticity.

Additionally, we consider a system model in which an authenticated classical channel is connected between the sender and receiver.
The classical channel must be authenticated because the proposed scheme must prevent man-in-the-middle attacks, such as spoofing~\cite{sa00}.
We also consider scenarios in which arbitrary quantum states, such as the intermediate results of quantum computing~\cite{dc21}, are sent once, rather than repeatedly transmitting the same quantum state.
To be precise, this "send-once" constraint applies to the specific combination of a state and the random keys that will be described in the following section~\ref{kg}.
While the same state $|\psi'\rangle$ cannot be resent using the same set of random keys, re-transmission is permitted with a newly generated set if a verification step fails.

\subsection{KeyGen}
\label{kg}
To transmit arbitrary quantum states, the sender first measures the quantum bit error rate of the quantum channel.
Based on this information, the sender determines the error correction capability $t$ of the QECCs and selects an $[[n, k, d]]$ QECC.
In this $[[n, k, d]]$ notation, $n$ is the number of physical qubits and $k$ is the number of logical qubits.
The code's minimum distance $d$ is selected to provide the required error correction capability $t$, according to the relation $t = \lfloor (d-1)/2 \rfloor$.
To facilitate the security assessment discussed in the APPENDIX, we consider only non-degenerate quantum codes.
A non-degenerate code  satisfies the condition that each correctable error yields linearly independent results when applied to elements of the code~\cite{qec04, qec07}.

Subsequently, the sender generates four bit-strings to create four encryption keys, $\mathbf{\kappa^1}$, $\mathbf{\kappa^2}$, $\mathbf{\kappa^3}$, and $\mathbf{\kappa^4}$.
The keys $\mathbf{\kappa^1}$, $\mathbf{\kappa^2}$, and $\mathbf{\kappa^3}$ are $(k-k')$-bit strings, each generated independently and uniformly at random.
The key $\mathbf{\kappa^4}$ is a $k$-bit string with a Hamming weight of $k'$, generated by selecting $k'$ bit positions uniformly at random and setting them to 1.
The roles of these keys are discussed in section~\ref{EE}.

\subsection{Encoding and Encryption}
\label{EE}
\subsubsection{State preparation}
The quantum state that the sender wishes to transmit is an arbitrary $k'$-qubit data $|\psi' \rangle = \sum_{i=0}^{2^{k'}-1} c_i |m_i \rangle$ where $|m_i\rangle$ denotes the computational basis states and $c_i$ are the complex amplitudes associated with the basis states.
To prevent an eavesdropper from intercepting the entire quantum state and subsequently transmitting spoofed data—that is, intercept-and-resend attacks—or from extracting information using ancilla states, unitary operations, and measurements, we randomly insert $k-k'$ dummy states $|D\rangle$ into the data sequence.
These dummy states are randomly chosen from two sets of mutually unbiased basis (MUB) states: $\{|0\rangle, |1\rangle \}$ and $\{|+\rangle, |-\rangle \}$.

At this point, the dummy states $|D\rangle $ are generated using $\mathbf{\kappa^1}$, $\mathbf{\kappa^2}$, and $\mathbf{\kappa^3}$ generated in section~\ref{kg}.
This can be expressed as
\begin{equation}
|D\rangle = \bigotimes_{i=1}^{k-k'} X^{\mathbf{\kappa^{1}_{i}}}Z^{\mathbf{\kappa^{2}_{i}}}H^{\mathbf{\kappa^{3}_{i}}} |0\rangle^{\otimes k-k'} = U_{MUB}|0\rangle^{\otimes k-k'},
\end{equation}
where $X$ and $Z$ are Pauli operators, $H$ is the Hadamard operator, and $X^{\mathbf{\kappa^{1}_{i}}}$, $Z^{\mathbf{\kappa^{2}_{i}}}$, and $H^{\mathbf{\kappa^{3}_{i}}}$ denote operators that apply the respective operation when the corresponding bit in $\kappa^{1}$, $\kappa^{2}$, or $\kappa^{3}$ is 1, and the identity otherwise.
For notational simplicity, we define the combined operator as $U_{MUB}$, which transforms the ancillary zero state into a mutually unbiased basis (MUB) state.

\subsubsection{State encoding}
The prepared states must be encoded using QECCs to transmit the quantum state reliably.
Thus, the $n$-qubit state $|\psi \rangle$ prepared by the sender for $[[n, k, d]]$ QECC encoding can be expressed as
\begin{align}
|\psi \rangle &=(P_{\mathbf{\kappa^4}}\otimes I^{\otimes n-k})|\psi' \rangle U_{MUB}|0\rangle^{\otimes k-k'} |0\rangle^{\otimes n-k}, \nonumber \\
              &= (P_{\mathbf{\kappa^4}}\otimes I^{\otimes n-k})(I^{\otimes k'}\otimes U_{MUB}\otimes I^{\otimes n-k}) |\psi' \rangle |0\rangle^{\otimes n-k'}, \nonumber \\
              &=U_c |\psi' \rangle |0\rangle^{\otimes n-k'},
\end{align}
where $P_{\mathbf{\kappa^4}}$ is the permutation operator, determined by the key $\mathbf{\kappa^4}$ described in section~\ref{kg}, that shuffles the $k'$ data and $k-k'$-qubit dummy states $|D\rangle$.
$U_c$ denotes the composite operation that combines the permutation $P_{\mathbf{\kappa^4}}$ with the MUB transformation $U_{MUB}$.

The sender encodes the state $|\psi \rangle$ into a logical state using the encoding operator $U_E$ associated with the selected QECC
\begin{equation}
|\psi\rangle_{L} = U_E|\psi\rangle.
\end{equation}

\subsubsection{State encryption}
To perform encryption, a random Pauli error operator $E_{un}$, which the chosen QECC cannot correct, is injected into the encoded logical state $|\psi \rangle_L$.
The uncorrectable error $E_{un}$ is injected into the transmitted state, preventing eavesdroppers or intermediate nodes from obtaining information about the state.
Additionally, after receiving the state, the receiver uses it to verify whether the received state is indeed the one sent by the sender, functioning as a signature.
These functionalities will be elaborated in section~\ref{dd} and the set of $E_{un}$ that the sender can choose is detailed in the APPENDIX.

The resulting encoded and encrypted state $|\psi \rangle_E$ is as follows:
\begin{equation}
|\psi \rangle_E = E_{un}|\psi\rangle_{L}.
\end{equation}
The sender calculates the syndrome $s$ of the injected $E_{un}$, then transmits the encoded state $|\psi \rangle_E$ through the quantum channel and transmits the syndrome $s$ through the classical channel to the receiver.
We assume that the quantum channel is noisy and insecure, while the classical channel is error-free and authenticated.

\subsection{Decryption}
\label{dd}

\subsubsection{Syndrome-based error correction before verification}
Upon receiving $|\psi \rangle_E$ and $s$, the receiver extracts the syndrome of $|\psi \rangle_E$ and performs error correction based on $s$.
Generally, error correction using syndromes is performed based on the all-zero syndrome to revert to an error-free state.
However, because the difference between syndromes represents the channel error that has occurred, error correction based on the syndrome $s$ can correct the channel error~\cite{sd00}.

\subsubsection{Verification of $E_{un}$}
Thereafter, the receiver transmits an $\textit{ACK}_1$ to inform the sender that $|\psi \rangle_E$ has been received, where $\textit{ACK}$ (acknowledgment) refers to a signal in data networking that confirms the successful receipt of a transmitted message~\cite{arq00}.
Upon receiving the $\textit{ACK}_1$, the sender transmits $E_{un}$ to the receiver via a classical channel.
The receiver applies $E_{un}$ to $|\psi \rangle_E$ and performs syndrome extraction again to verify if an all-zero syndrome is obtained.
If the syndrome is not all-zero, the receiver assumes a potential eavesdropping attack and aborts the process.

\subsubsection{Dummy states validation}
If an all-zero syndrome is extracted, the receiver transmits an $\textit{ACK}_2$ to the sender again indicating the extraction of an all-zero syndrome.
Upon receiving the $\textit{ACK}_2$, the sender transmits a random permutation operator $P_{\kappa^4}$ and $U_{MUB}$ through the classical channel.
Subsequently, the receiver measures each mutually unbiased state in the corresponding $U_{MUB}$ to verify the consistency between the results encoded by $U_{MUB}$ and the measurement outcomes.
If errors are detected, it is assumed that an eavesdropper attempts to extract information, and the process is aborted.

\subsection{Distance extension}
\begin{figure}[ht]
\centerline{\includegraphics[width=.65\linewidth]{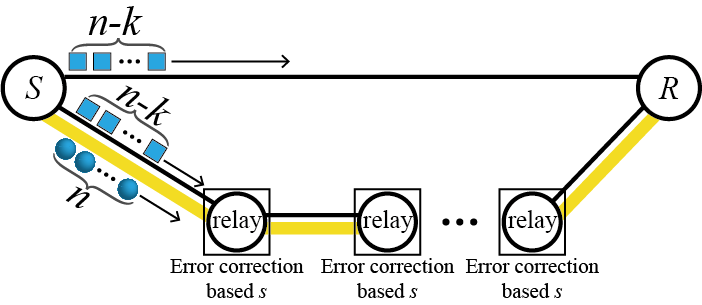}}
\caption{\textbf{Distance extension of the proposed scheme} Relay nodes can perform error correction based on $s$, ensuring that relay nodes cannot obtain any information about the quantum states.}
\label{de}
\end{figure}
The advantage of the proposed approach lies in its ability to extend distance despite encryption, as error correction is still feasible.
As illustrated in figure~\ref{de}, similar to the same process that the receiver performs in Sect.~\ref{dd}, relay nodes perform error correction based on $s$, and thereafter pass it to the next node, enabling fault-tolerant transmission.
Furthermore, because $E_{un}$ is injected, relay nodes cannot obtain any information regarding the quantum states through syndrome extraction~\cite{qec07}.

\section{Security analysis}
In this section, we analyze the security of the proposed scheme.
The only operators that do not affect the syndrome of a quantum state encoded with a quantum error-correcting code are stabilizers and logical operators~\cite{qec07}.
Additionally, the dummy states $|D\rangle$ introduced during the state preparation in the proposed scheme can be modified by logical operators.
Therefore, if an eavesdropper attempts to extract information using a specific operation or a measurement involving logical ancilla states, the syndrome $s$ of $E_{un}$ or $|D\rangle$ may become corrupted, making such attempts detectable.
In other words, even if Eve is assumed to be a computationally unbounded quantum adversary~\cite{eve}, the inability of stabilizers to reveal logical information ensures that no data can be extracted without disturbing the syndrome or the dummy states.
The only case where neither is affected is an attack using stabilizers, but it is well known that such attacks cannot extract information about the quantum state~\cite{qec07}.
Therefore, to demonstrate that such forms of attacks can be detected, we first examine countermeasures against \textit{intercept-and-resend attacks}, which serve as an example where the syndrome $s$ alone cannot detect the attack, but $|D\rangle$ enables detection.
Additionally, we analyze the accessible information in the proposed scheme to evaluate its security in terms of potential information leakage~\cite{ai00, ai01, ai02}.
We demonstrate that as the length of the message $k'$ increases, the potential information leakage can be sufficiently minimized.

\subsection{Intercept-and-resend attacks}
Intercept-and-resend attacks involve an eavesdropper intercepting the transmitted quantum state $|\psi \rangle_E$ and sending a spoofed quantum state with a spoofed injected error $E^{'}_{un}$ that matches the syndrome $s$ to the receiver, masquerading as the sender.
Thereafter, the eavesdropper intercepts the $E_{un}$ sent by the legitimate sender to extract information from $|\psi \rangle_E$.

This attack can be detected because when the receiver applies the received $E_{un}$ to the spoofed quantum state of the eavesdropper, the difference between $E_{un}$ and $E^{'}_{un}$ results in a non-zero syndrome during the subsequent syndrome extraction.
In this case, it could be problematic if the eavesdropper retains the information of $|\psi' \rangle$ using the intercepted quantum state and $E_{un}$.

To prevent this, dummy states $|D\rangle$ defined in MUBs are mixed with $|\psi'\rangle$, and this information is only disclosed when the sender receives the $\textit{ACK}_1$ indicating an all-zero syndrome.
This permutation-based security relies on $P_{\kappa^4}$, that is, the number of all possible combinations of the key $\kappa^4$, denoted as $|\kappa^4|$~\cite{ps00, ps01}.
In the proposed scheme, the probability that an eavesdropper can extract $|\psi'\rangle$ is ${k\choose k'}^{-1}$.

\subsection{Accessible information available to the eavesdropper}
To derive accessible information, we adopt the perspective of the eavesdropper.
Since the eavesdropper knows neither the arbitrary quantum data being sent nor the random keys used for encryption, the state is indistinguishable from a uniform mixture of all possible data.
This modeling clarifies that the security relies on the randomness of the encryption keys, not on any assumption that the message itself must be random.
In addition, because of the influence of the data and randomly permuted dummy states $|D\rangle$ defined in MUBs, $E_{un}$ appears as a state in which all possible Pauli error patterns are mixed.
Thus, when considering only the data, uncorrectable error applied to the state differs from the actual injected $E_{un}$.
This can be mathematically verified as
\begin{align}
|\psi \rangle_E &= E_{un}U_E U_c |\psi' \rangle |0\rangle^{\otimes n-k'}, \nonumber \\
%                &= E_{un}U_E U_c U^{-1}_E U_E |\psi' \rangle |0\rangle^{\otimes n-k'} \nonumber \\
                &= E_{un} U^{'}_c U_E |\psi' \rangle |0\rangle^{\otimes n-k'},
                \label{sec}
\end{align}
where $U^{'}_c=U_E U_c U^{-1}_E$.
Thus, although the eavesdropper can narrow down the candidate $E_{un}$ based on the syndrome information $s$, they must identify $E_{un} U^{'}_c$, which necessitates the consideration of all possible Pauli error patterns~\cite{ps00, ps01}.

Consequently, the description of the quantum state after the receiver's error correction process perceived by the eavesdropper, $\rho_{E}$,
\begin{equation}
\rho_{E} = \frac{1}{2^{k'}} \frac{1}{4^n} \sum_{i=0}^{2^{k'}-1}\sum_{\mathbf{j},\mathbf{k}\in \{0,1\}^{n}} X^{\mathbf{j}}Z^{\mathbf{k}}|\psi \rangle_{L,i} \langle \psi |_{L,i}Z^{\mathbf{k}}X^{\mathbf{j}},
\end{equation}
where $|\psi \rangle_{L,i}$ denotes the logical basis for $|m_i\rangle$, and the vectors $\mathbf{j}, \mathbf{k} \in \{0,1\}^n$ are $n$-bit binary strings that specify the locations of the Pauli $X$ and $Z$ operators, respectively.

The accessible information is defined as the maximum mutual information, $I_{acc}(M;E)$,
\begin{equation}
I_{acc}(M;E) = \max_{\Lambda} I(M;Y).
\end{equation}
where $M$ is the message of the sender, $E$ is system of the eavesdropper, and $Y$ is a random variable obtained from the measurements $\Lambda$ of the eavesdropper.
According to the convexity of mutual information, the maximum can be achieved through a positive operator-valued measure, $\{\Lambda_y\}$ with rank-one elements, such that $\Lambda_y \geq 0$, $\sum_{y} \Lambda_y = \mathds{I}$~\cite{ai01, povm00},
\begin{equation}
\Lambda_y=\mu_y |\phi_y \rangle \langle \phi_y|,
\end{equation}
where $|\phi_y \rangle$ are unit vectors and $\mu_y$ are positive numbers such that $\sum_{y} \mu_y=2^n$.

The measurement results follow the probability distribution:
\begin{equation}
p_{Y}(y)=\mu_y \langle \phi_y|\rho_{E}|\phi_y \rangle.
\label{pd1}
\end{equation}
For a given $m$, one of the entire basis of $|\psi'\rangle$, the conditional probability of a measurement outcome is
\begin{equation}
p_{Y|M=m}(y)=\mu_y \langle \phi_y|\rho^{m}_{E}|\phi_y \rangle.
\label{pd2}
\end{equation}
with
\begin{equation}
\rho^{m}_{E}=\frac{1}{4^n} \sum_{\mathbf{j},\mathbf{k}\in \{0,1\}^{n}}X^{\mathbf{j}}Z^{\mathbf{k}}|\psi \rangle_{L} \langle \psi |_{L}Z^{\mathbf{k}}X^{\mathbf{j}}.
\end{equation}
The accessible information is given by
\begin{align}
I_{acc}(M;E) = &\max_{\Lambda} \{ -\sum_{y} p_{Y}(y)\log p_{Y}(y)  \nonumber\\
                  &+ \frac{1}{2^{k'}}\sum_{y,m} p_{Y|M=m}(y)\log p_{Y|M=m}(y) \}, \nonumber\\
             = &\max_{\Lambda} \sum_{y} \mu_y \{ -\langle \phi_y|\rho_{E}|\phi_y \rangle\log \langle
             \phi_y|\rho_{E}|\phi_y \rangle  \nonumber\\
                  &+ \frac{1}{2^{k'}}\sum_{m} \langle \phi_y|\rho^{m}_{E}|\phi_y \rangle\log \langle \phi_y|\rho^{m}_{E}|\phi_y \rangle \},
                  \label{ia}
\end{align}
where the term $\mu_y$ inside the logarithm in equation~(\ref{ia}) can be canceled out using the relationship between equations~(\ref{pd1}) and~(\ref{pd2}).

From the perspective of an eavesdropper, errors are perceived as maximally mixed.
Therefore, the higher the effective code rate, $R_{\mathrm{eff}} = k'/n$---which can differ from the code rate $R=k/n$ due to the presence of $k-k'$ dummy qubits---the closer $\rho_E$ and $\rho^{m}_E$ approach maximally mixed states (MMSs).

By applying the results in~\cite{ai02} with the matrix Chernoff bound and Maurer bound,
\begin{align}
\langle \phi|\rho_{E}|\phi \rangle &\leq (1+\epsilon)2^{-n}, \label{lb}\\
\langle \phi|\rho^{m}_{E}|\phi \rangle &\geq (1-\epsilon)2^{-n}. \label{ub}
\end{align}
Then, by substituting equation~(\ref{lb}) and~(\ref{ub}) into equation~(\ref{ia}), the accessible information can be obtained as
\begin{equation}
I_{acc}(M;E) \leq 2\epsilon n,
\end{equation}
where $\epsilon>0$.
As $k'$  and $n$ increase, $\rho_{E}$ and $\rho^{m}_{E}$ asymptotically approach MMSs, ensuring that any accessible information for an eavesdropper becomes increasingly randomized.
This strengthens security by making the extracted information less distinguishable from noise.
However, since the total message length scales with $n$, the absolute amount of leaked information also increases.
This is an inherent effect of encoding a larger message rather than a weakness of the scheme, as the fraction of leaked information relative to the total message content continues to decrease.
Therefore, the quantum state can be transmitted using the proposed scheme while ensuring that the information leaked to the eavesdropper remains sufficiently negligible.

\section{Resource and fidelity analysis}
\label{eval}
In this section, we present a comparative analysis of the resource overhead and fidelity between a long-distance entanglement distribution (LDED) scheme and the proposed scheme.
We acknowledge that both channel errors and qubit loss are major sources of error in long-distance quantum communication.
The proposed scheme, based on QECCs, is capable of correcting both types of errors.
A general $[[n,k,d]]$ code can correct up to $t = \lfloor (d-1)/2 \rfloor$ Pauli errors with unknown locations. 
Alternatively, it can correct up to $d-1$ erasure errors (losses) whose locations are known~\cite{ee00}.
For a channel with both, it can correct a combination of $t$ Pauli errors and $r$ erasures provided that $2t+r<d$~\cite{ee01}.
However, from a security standpoint, the most relevant threat model involves an eavesdropper's attack, which more closely resembles a channel characterized by Pauli errors.
For this reason, our primary comparative analysis focuses on the depolarizing channel.
A specific analysis of our scheme's logical error rate in a channel with both depolarizing errors and qubit loss is provided in the \hyperref[secB]{Appendix B}.

\subsection{Resource overhead modeling}
This section presents a modeling-based analysis of the resource overhead incurred by the LDED, focusing on entanglement swapping and purification over a linear chain network.
We subsequently apply the derived overhead to evaluate the performance of the proposed scheme in terms of fidelity under comparable resource conditions.
We consider the LDED over a linear chain network consisting of $2^N + 1$ nodes.
This architecture requires entanglement swapping and purification to maintain high fidelity.

Let $N_{EP}$ denote the number of ancillary qubits required for entanglement purification. Assuming all entanglement swapping operations succeed, the purification overhead is given by:
\begin{equation}
N_{EP} = \sum_{i=0}^N N_A 2^{N-i} \times 2,
\end{equation}
where $N_A$ is the number of ancilla qubits for a single purification step.

The number of BSMs required for entanglement swapping is:
\begin{equation}
O_{ES} = \sum_{i=1}^N 2^{N-i}.
\end{equation}

To increase the BSM success probability, $P_{ES}=1-1/2^{n_{bsm}}$, we employ logical BSMs.
The probability that all $O_{ES}$ BSMs succeed is $(1-1/2^{n_{bsm}})^{O_{ES}}$, and the number of repetitions required for at least one successful attempt is:
\begin{equation}
\left\lceil \left(\frac{2^{n_{bsm}}}{2^{n_{bsm}}-1}\right)^{O_{ES}} \right\rceil.
\end{equation}

The total number of qubits $N_T$ required for one successful LDED is:
\begin{equation}
N_T = (N_{EP} + 2O_{ES}(n_{bsm} - 1)) \times \left\lceil \left(\frac{2^{n_{bsm}}}{2^{n_{bsm}}-1}\right)^{O_{ES}} \right\rceil.
\end{equation}

\subsection{Fidelity estimation}
\label{fe}
Using the resource overhead analyzed in the previous section, we evaluate the transmission fidelity achieved by the proposed scheme and compare it with that of the LDED involving entanglement swapping and purification.

We consider a noise model in which only depolarizing noise is applied at the final purification stage.
All preceding steps in the LDED scheme—such as entanglement generation, entanglement swapping, intermediate purification, and state preparation—are assumed to be noiseless.
Likewise, gate operations and measurements in both schemes are also treated as ideal.
These assumptions are introduced to enable a direct comparison between the two schemes under a controlled setting.

For the proposed scheme, the fidelity is approximated based on the logical error probability $p_L$ of a quantum stabilizer code.
Under a depolarizing channel with physical error rate $p$, the logical error probability is given by~\cite{le00}:
\begin{equation}
p_L = 1 - \sum_{i=0}^t \binom{N_T}{i} p^i (1-p)^{N_T-i}.
\end{equation}
To estimate $t$, we apply the quantum Singleton bound~\cite{sb00},
\begin{equation}
N_T - k \geq 2(d - 1),
\end{equation}
which we reformulate as:
\begin{equation}
\frac{N_T - k}{4} \geq t.
\end{equation}

Assuming a code rate \( R = 1/2 \), the lower bound on $p_L$ becomes:
\begin{equation}
p_L \geq 1 - \sum_{i=0}^{\lfloor N_T/8 \rfloor} \binom{N_T}{i} p^i (1-p)^{N_T-i},
\end{equation}
and the fidelity of the proposed scheme is accordingly approximated as:
\begin{equation}
F_{\mathrm{our}} \approx \sum_{i=0}^{\lfloor N_T/8 \rfloor} \binom{N_T}{i} p^i (1-p)^{N_T-i}.
\label{four}
\end{equation}

For the LDED, assuming that the initial Bell state is prepared as the $|\Phi^+\rangle$ state and each qubit is independently subject to depolarizing noise, the resulting initial fidelity is given by $F_{\mathrm{ini}} = (1 - p)^2 + \frac{p^2}{3}$, where $p$ denotes the physical error rate.
The fidelity expression is derived based on the model in Ref.~\cite{ep06}, which demonstrates high purification performance even with a small number of ancilla qubits, and is given by:
\begin{equation}
F_{\mathrm{LDED}}= \frac{(F_{\mathrm{ini}})^{N_A+1}}{(F_{\mathrm{ini}})^{N_A+1} + (1 - F_{\mathrm{ini}})^{N_A+1}}.
\label{flded}
\end{equation}

\subsection{Comparative performance analysis}
\label{ce}

\begin{figure}[t]
    \centering
    \includegraphics[width=0.8\linewidth]{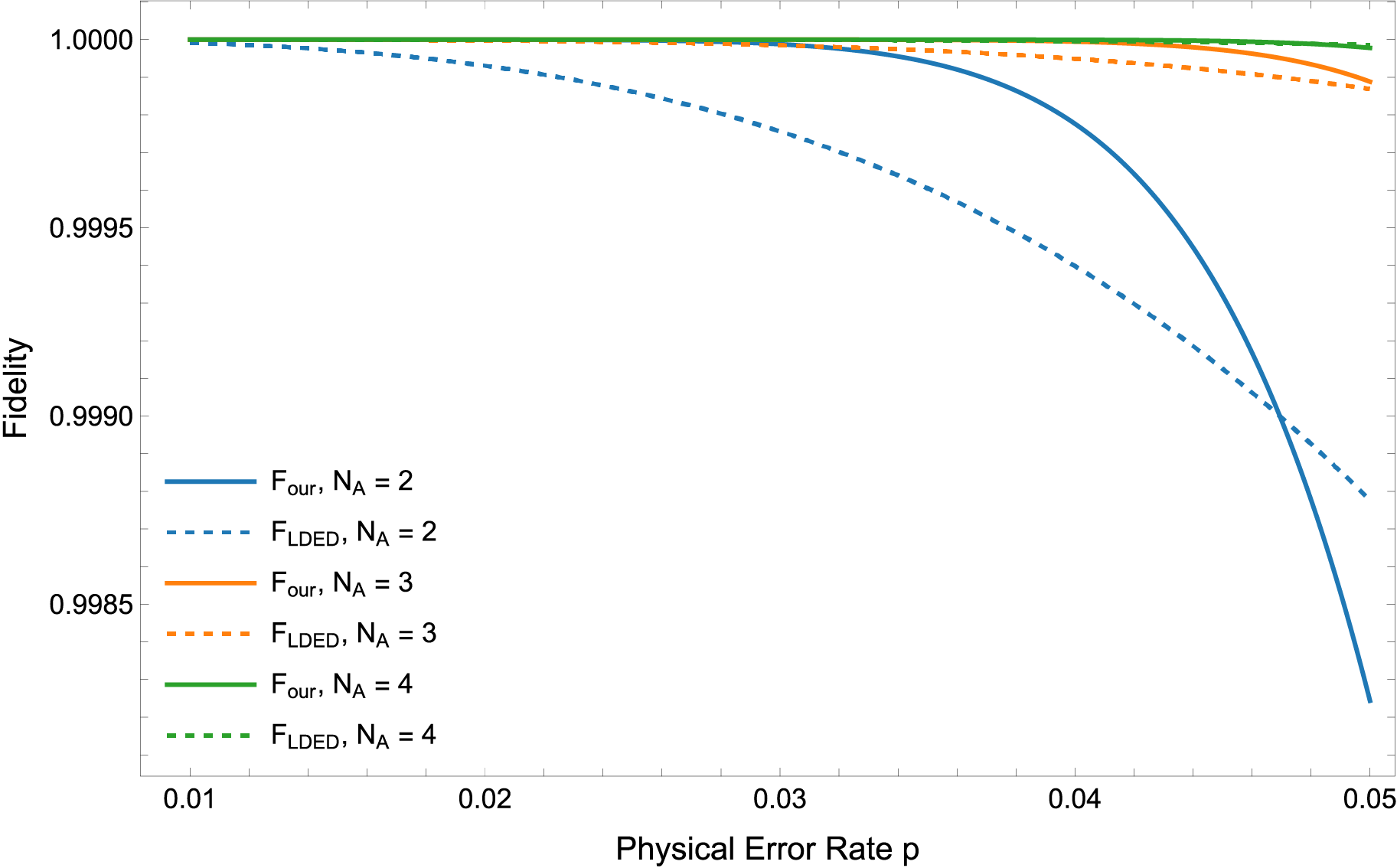}
    \caption{Fidelity comparison between $F_{\mathrm{our}}$ and $F_{\mathrm{LDED}}$ with varying purification resource levels ($N_A = 2$, $3$, and $4$).
    The network consists of five nodes (i.e., $N=2$), and the Bell state measurement parameter is $n_{\mathrm{bsm}} = 2$.
    Solid lines indicate $F_{\mathrm{our}}$, and dashed lines indicate $F_{\mathrm{LDED}}$.
    The results show that in the low-error regime, the proposed scheme achieves higher fidelity.}
    \label{fig:fpc}
\end{figure}

To provide a practical and quantitative comparison between the proposed scheme and the LDED scheme, we evaluate their fidelity performance using the analytic models derived in Section~\ref{fe}.
We assume a network consisting of five nodes (i.e., $N=2$), and consider LDED configurations with $N_A = 2$, $3$, and $4$ ancillary Bell states per end node.
The fidelity of each configuration is computed using the closed-form expressions for $F_{\mathrm{our}}$ and $F_{\mathrm{LDED}}$, and the resulting fidelity trends across different physical error rates $p$ are shown in figure~\ref{fig:fpc}.

\begin{figure}[t]
    \centering
    \includegraphics[width=\linewidth]{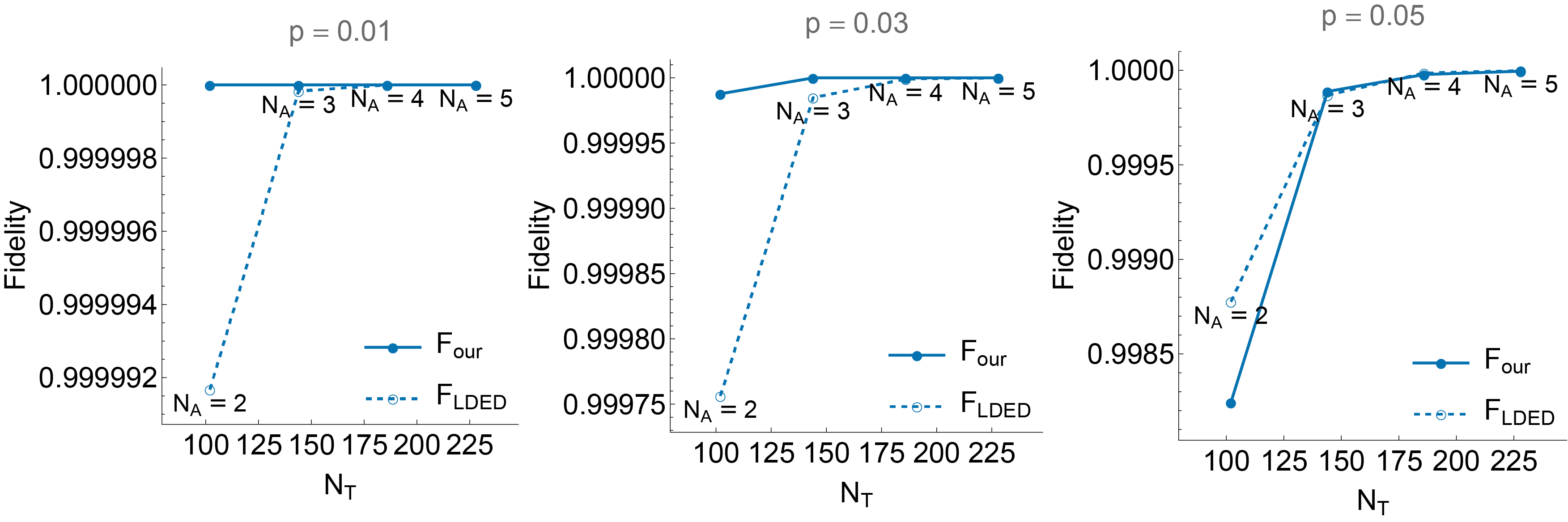}
    \caption{Fidelity as a function of total qubit overhead ($N_T$ with $N_A = 2, 3, 4, 5$) for $F_{\mathrm{our}}$ and $F_{\mathrm{LDED}}$ under various physical error rates ($p = 0.01$, $0.03$, and $0.05$).
    The results illustrate that the proposed scheme achieves high fidelity with lower overhead in low-noise regimes.}
    \label{fig:oc}
\end{figure}

To further investigate the trade-off between fidelity and total qubit overhead, we fix the physical error rate $p$ and evaluate how the fidelity scales with $N_T$.
Figure~\ref{fig:oc} presents the results for three representative values of $p$ ($0.01$, $0.03$, and $0.05$), illustrating the behavior of both schemes across different noise levels.
At lower error rates (e.g., $p = 0.01$ or $0.03$), the proposed scheme achieves high fidelity with substantially fewer qubits than LDED.
As $p$ increases, the required overhead to maintain comparable fidelity also increases, narrowing the performance gap between the two schemes.
These results indicate that the proposed approach is particularly advantageous in low-noise regimes, offering strong fidelity with reduced resource requirements.

\begin{table}[htb!]
\centering
\caption{Comparison of fidelity and resource overhead between the proposed scheme and the LDED under $n_{\mathrm{bsm}} = 2$ and different parameter settings.
``Enhanced efficiency'' refers to configurations where the proposed scheme achieves slightly higher fidelity than the LDED while using the minimum number of qubits.
``Resource reduction'' denotes the percentage of qubit savings achieved relative to the LDED baseline.}
\label{tab:ex}
\begin{tabular}{|c|c|c|c|c|c|}
\hline
\textbf{Scheme} & $p$ & $N$ & $N_T$ & \textbf{Fidelity} & \makecell{\textbf{Resource} \\ \textbf{reduction (\%)}} \\
\hline
\textbf{LDED} (baseline) &  &  & 102 & $1 - 8.33 \times 10^{-6}$  &  \\
\textbf{Proposed} (matched resource)& 0.01 & 2 & 102 & $1 - 4.08 \times 10^{-11}$ & 60.78 \\
\textbf{Proposed} (enhanced efficiency) &  &  & 40  & $1 - 2.87 \times 10^{-6}$  &  \\
\hline
\textbf{LDED} (baseline) &  &  & 779 & $1 - 2.85 \times 10^{-6}$  &  \\
\textbf{Proposed} (matched resource)& 0.02 & 3 & 779 & $1 - 1.32 \times 10^{-14}$ & 90.76 \\
\textbf{Proposed} (enhanced efficiency) &  &  & 72  & $1 - 1.77 \times 10^{-6}$  &  \\
\hline
\textbf{LDED} (baseline)&  &  & 113 & $1 - 1.52 \times 10^{-5}$  &  \\
\textbf{Proposed} (matched resource)& 0.03 & 2 & 113 & $1 - 1.61 \times 10^{-6}$  & 22.12 \\
\textbf{Proposed} (enhanced efficiency) &  &  & 88  & $1 - 1.31 \times 10^{-5}$  &  \\
\hline
\textbf{LDED} (baseline)&  &  & 1003 & $1 - 9.52 \times 10^{-7}$  &  \\
\textbf{Proposed} (matched resource)& 0.03 & 3 & 1003 & $1 - 2.81 \times 10^{-14}$ & 88.04 \\
\textbf{Proposed} (enhanced efficiency) &  &  & 120  & $1 - 6.92 \times 10^{-7}$  &  \\
\hline
\end{tabular}
\end{table}

As a simple example, consider the case where $N_A = 2$, $p = 0.01$, $n_{\mathrm{bsm}} = 2$, and the total number of nodes is 5.
In this setting, the number of qubits required for entanglement purification is $N_{\mathrm{EP}} = 28$, resulting in a total resource count of $N_T = 102$. The fidelity achieved by our scheme is $F_{\mathrm{our}} \approx 1 - 4.08 \times 10^{-11}$, while the LDED achieves $F_{\mathrm{LDED}} \approx 1 - 8.33 \times 10^{-6}$.
Notably, even with just $N_T = 40$ qubits, the proposed scheme achieves $F_{\mathrm{our}} \approx 1 - 2.87 \times 10^{-6}$, surpassing the LDED in fidelity.
Table~\ref{tab:ex} summarizes a set of representative scenarios comparing the proposed scheme and the LDED under various values of $p$, $N_A$ and network size $N$.
In each case, the proposed method achieves comparable or superior fidelity while requiring significantly fewer physical qubits.
The last column quantifies this advantage, showing that our scheme can reduce qubit requirements by more than 60\% to 90\% in low-error regimes, with minimal or even improved fidelity loss.
Such results highlight the resource-efficiency of our approach, requiring substantially fewer physical qubits in low-error regimes.

\section{Conclusion}
In this study, we proposed a novel fault-tolerant and secure quantum communication scheme leveraging uncorrectable error injection.
Unlike conventional quantum teleportation-based approaches, which require entanglement distribution, entanglement swapping, and purification, our method eliminates the need for pre-distributed entanglement while ensuring secure and fault-tolerant quantum state transmission.
By encoding quantum states with QECCs and introducing uncorrectable errors, we enhanced both the fault tolerance and security of the transmission process while reducing the resource overhead associated with entanglement management.

Our security analysis demonstrated that the proposed scheme is resilient against intercept-and-resend attacks, as unauthorized modifications to the transmitted state can be detected through syndrome extraction and verification.
Furthermore, the presence of uncorrectable errors prevents an eavesdropper from extracting meaningful information from an intercepted state.
The comparative performance analysis confirmed that the proposed scheme effectively reduces the overhead associated with quantum state transmission while maintaining high fidelity in low physical error regimes.
By eliminating the complexities of entanglement distribution and minimizing the number of required quantum operations, our approach provides a resource-efficient solution for long-distance quantum communication. These findings indicate that the proposed approach offers a scalable and practical alternative to conventional quantum state transmission methods, particularly in large-scale quantum networks.

Future research could focus on experimentally validating the proposed scheme using near-term quantum hardware or by employing existing quantum network simulation tools, as well as integrating it into emerging quantum communication frameworks, including quantum key distribution and quantum repeaters.
While the current security analysis addresses representative attack scenarios, extending the evaluation to include more sophisticated quantum adversaries will further strengthen the protocol's robustness.
In addition, exploring optimization strategies for encoding efficiency and analyzing the trade-offs between security and fidelity across diverse quantum network configurations will be valuable for enhancing practical deployment.
A comprehensive comparative analysis against LDED schemes including qubit loss also remains a crucial future work. 
Moreover, investigating the interplay between our scheme and high-throughput frameworks, such as multicarrier CV-QKD systems, presents a promising research avenue~\cite{mc00, mc01}.
By further refining the method and broadening its applicability, this research can contribute to the advancement of secure, scalable, and fault-tolerant quantum networking technologies.

\bmhead{Acknowledgements}
This research was supported by Korea Institute of Science and Technology Information (KISTI). (No. K25L5M2C2). This research was supported by the National Research Council of Science \& Technology (NST) grant by the Korea government (MSIT) (No. CAP22053-000). K.J. acknowledges support by the National Research Foundation of Korea (NRF) through a grant funded by the Ministry of Science and ICT (Grant No. RS-2025-00515537), the Institute for Information \& Communications Technology Promotion (IITP) grant funded by the Korean government (MSIP) (Grant No. RS-2025-02304540), and the National Research Council of Science \& Technology (NST) (Grant No. GTL25011-401).

\section*{Declarations}
\begin{itemize}
\item Funding: Not applicable.
\item Data availability: Not applicable.
\item Materials availability: Not applicable.
\item Code availability: Not applicable.
\item Conflict of interest/Competing interests (check journal-specific guidelines for which heading to use): On behalf of all authors, the corresponding author states that there is no conflict of interest.
\item Ethics approval and consent to participate: This work poses no ethical issues or challenges and is rightfully in line with the format
for writing manuscripts or articles. All authors consent to participate in this research or paper.
\item Consent for publication:  All authors consent to publish this research or paper.
\item Author contribution:  All authors contributed equally to the paper.
\end{itemize}

\begin{appendices}

\section{Analysis of number of uncorrectable errors}
\label{dis}
In this section, we discuss the number of uncorrectable errors for encryption.
The security of the proposed scheme is primarily determined by the number of uncorrectable errors assigned to each syndrome.
The number of uncorrectable errors assigned to a syndrome, $N_u$, can be estimated as follows:
\begin{equation}
N_u \sim \frac{4^n-\sum_{i=0}^{t}3^i \binom{n}{i}(2^{n-k}+2^{2k})}{2^{n-k}} \times \frac{1}{2^{n-k}}.
\label{Nu}
\end{equation}
where $4^n$ of equation~(\ref{Nu}) represents the number of all Pauli error patterns of length $n$, and $\sum_{i=0}^{t}3^i \binom{n}{i}$ denotes the number of all errors that the QECCs can correct.
For simplicity, we refer to this as $N_c$.
The subsequent terms multiplied by $N_c$ represents the number of errors with different weights sharing the same syndrome as correctable errors within $N_c$.
Between these terms, $2^{n-k}$ represents the total number of stabilizers.
When multiplied by $N_c$, it represents errors that share the same syndrome and behavior as correctable errors within $N_c$.
The value $2^{2k}$ represents the total number of logical Pauli operators.
When it is multiplied by $N_c$, it represents errors that share the same syndrome but have different behaviors because of logical operators, thereby being uncorrectable.
In other words, the numerator represents the total number of errors associated with uncorrectable syndromes.
Errors with uncorrectable syndromes, even when multiplied by stabilizers, exhibit the same behavior and syndrome.
Therefore, they should be considered as single errors.
To account for this, we adjusted by dividing by the total number of stabilizers, $2^{n-k}$, which served as the denominator.
The final $\frac{1}{2^{n-k}}$ term was used to calculate the average number of uncorrectable errors allocated to each syndrome.
Here, $2^{n-k}$ represents the total number of syndromes which are bit strings of length $n-k$.
Using this information, the approximate number of completely different errors, denoted as $N_u$, that share the same syndrome as any given uncorrectable error could be determined.
$N_c$ was substituted with a term containing $n$ and $k$ using the Hamming bound for QECCs~\cite{qh00}.
The quantum Hamming bound is expressed as
\begin{align}
    n-k &\geq \log\sum_{i=0}^{t}3^i \binom{n}{i}, \nonumber \\
2^{n-k} &\geq \sum_{i=0}^{t}3^i \binom{n}{i},
\end{align}
Rearranging equation~(\ref{Nu}) yields
\begin{align}
N_u &\geq \frac{4^n-2^{n-k}(2^{n-k}+2^2k)}{2^{n-k}} \times \frac{1}{2^{n-k}} \nonumber \\
    &\geq 2^{2k}-2^{n-k}(2^{k-n}+2^{4k-2n}).
\label{Nu2}
\end{align}
By substituting the code rate $k/n=R$, equation~(\ref{Nu2}) can be revised as,
\begin{equation}
N_u \geq 2^{2Rn} (1-(\frac{1}{2})^{(1-R)n})-1.
\label{Nu3}
\end{equation}
When $n$ is sufficiently large, equation~(\ref{Nu3}) can be approximated as follows:
\begin{equation}
N_u = \begin{cases}
\lfloor{2^{2Rn}-1}\rfloor, &\text{if } 2^{2Rn} \geq 1, \\
0, &\text{if } 2^{2Rn} < 1.
\end{cases}
\label{Nu4}
\end{equation}
\begin{figure}[t!]
\centerline{\includegraphics[width=.65\linewidth]{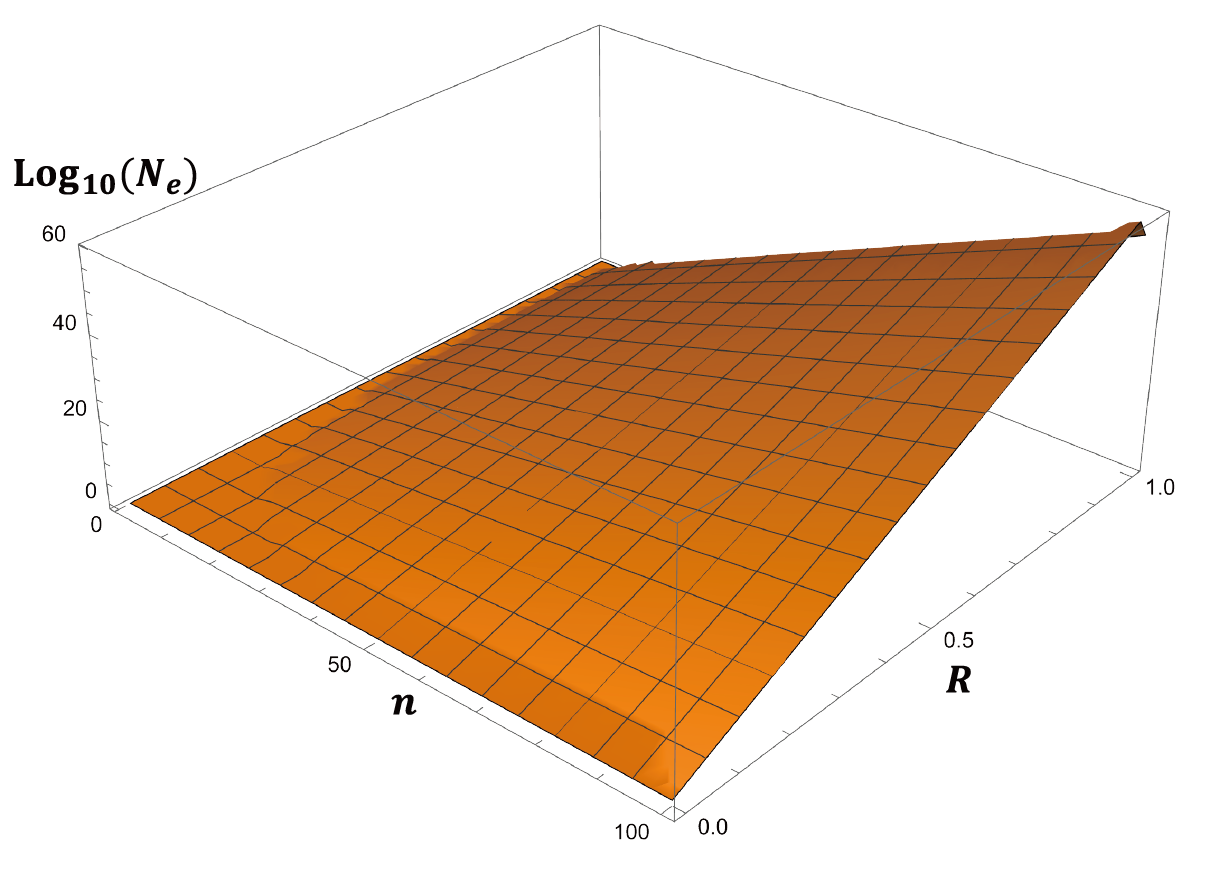}}
\caption{\textbf{Security analysis graph} This graph illustrates the order of the average number of distinct uncorrectable errors $N_u$ with the same syndrome as equation~(\ref{Nu4}).}
\label{fig:secg}
\end{figure}
As expressed in equation~(\ref{Nu4}), the graph of $N_u$ over the range $1\leq n\leq100$ is shown in figure~\ref{fig:secg}.
As shown in figure~\ref{fig:secg}, $N_u$ has an extremely small value when $R$ is less than $0.18$.
To ensure adequate security, QECCs with sufficiently large $n$, $R$ should be used.
As mentioned in section~\ref{ce}, for the case with a total of $5$ nodes and $102$ required qubits, the number of $N_u$ when $R = 0.5$ is approximately $5.0706\times 10^{30}$.

\section{Analysis of logical error rate with errors and qubit losses}
In this section, we discuss the logical error rate for the proposed scheme under a channel with both Pauli errors and qubit losses, as mentioned in section~\ref{eval}.
Let the physical error rate be denoted by $p$ and the probability of a qubit loss (erasure) by $l$.
Then the logical error rate, $P_L$, for an $[[n,k,d]]$ code under these conditions is given by:
\begin{align}
p_L &= \sum_{r=0}^{n}\sum_{t = t_{\min}(r)}^{n-r} \binom{n}{r}\binom{n-r}{t}l^{r} (1 - l)^{n-r}
( \frac{p}{1 - l})^{t}( 1 - \frac{p}{1 - l})^{n-r-t},\nonumber \\
&= \sum_{r=0}^{n}\sum_{t = t_{\min}(r)}^{n-r} \binom{n}{r}\binom{n-r}{t}l^{r}p^{t}(1-l-p)^{n-r-t},
\end{align}
where the minimum number of Pauli errors required to cause a logical error, $t_{\min}(r)$, depends on the number of qubit losses $r$ and is given by $t_{\min}(r) = \max\left(0,\left\lceil\frac{d - r}{2}\right\rceil\right)$.
This formula calculates the total logical error rate by summing the probabilities of all possible uncorrectable error events.
The summation over $r$ considers all possible numbers of qubit losses, from 0 to $n$.
For a fixed number of losses $r$, the term $\binom{n}{r} l^r (1-l)^{n-r}$ gives the binomial probability of exactly $r$ losses occurring.
The summation over $t$ then computes the probability of an uncorrectable error on the remaining $n-r$ qubits.
An uncorrectable error occurs if the number of Pauli errors $t$ meets or exceeds the required threshold $t_{\min}(r)$.
This probability is given by the second binomial term, which uses the conditional probability $p/(1-l)$ for a Pauli error to occur on a qubit given that it was not erased.

\end{appendices}

\end{document}